\documentclass[a4paper]{article}

\usepackage[pages=all, color=black, position={current page.south}, placement=bottom, scale=1, opacity=1, vshift=5mm]{background}

\usepackage[margin=1in]{geometry} 

\usepackage{amsmath}
\usepackage{amsthm}
\usepackage{amssymb}

\usepackage[utf8]{inputenc}
\usepackage{hyperref}
\hypersetup{colorlinks,allcolors=black}

\usepackage[sort&compress,numbers,square]{natbib}
\theoremstyle{plain}

\theoremstyle{definition}

\usepackage{graphicx, color}
\graphicspath{{fig/}}

\usepackage{algorithm, algpseudocode} 
\usepackage{mathrsfs} 

\usepackage{lipsum}

\title{consexpressionR: an R package for consensus differential gene expression analysis.} 

\author{Juliana Costa-Silva $^{1}$ \and David Menotti$^{1}$ \and Fabricio M. Lopes$^{2}$ \\ \and 
$^{1}$Department of Informatics, Federal University of  Paraná,\\ Rua Coronel Francisco Heráclito dos Santos, 100, 81531-990, Paraná, Brazil\\ \and  $^{2}$Department of Computer Science, Bioinformatics Graduate Program,\\ Federal University of Technology - Paraná,\\ Av. Alberto Carazzai, 1640 - Cornélio Procópio, Postal code: 86300-000, Paraná, Brazil.} 

\begin{document}
	\maketitle

\begin{abstract}
		\textbf{Motivation:} Bulk RNA-Seq is a widely used method for studying gene expression across a variety of contexts. The significance of RNA-Seq studies has grown with the advent of high-throughput sequencing technologies. Computational methods have been developed for each stage of the identification of differentially expressed genes. Nevertheless, there are few studies exploring the association between different types of methods. In this study, we evaluated the impact of the association of methodologies in the results of differential expression analysis. By adopting two data sets with qPCR data (to gold-standard reference), seven methods were implemented and assessed in R packages (EBSeq, edgeR, DESeq2, limma, SAMseq, NOISeq, and Knowseq), which was performed and assessed separately and in association. The results were evaluated considering the adopted qPCR data. \\
\textbf{Results:} Here, we introduce consexpressionR, an R package that automates differential expression analysis using consensus of at least seven methodologies, producing more assertive results with a significant reduction in false positives.\\
\textbf{Availability:} consexpressionR is an R package available via 
Source code and support are available at GitHub (https://github.com/costasilvati/consexpressionR). \\
\textbf{Contact:} \href{julianacostasilvati@gmail.com}{julianacostasilvati@gmail.com}\\
\textbf{Supplementary information:} Supplementary data are available at github.
		
		\noindent\textbf{Keywords:} ConsexpressionR, gene expression, RNA-Seq
	\end{abstract}
\tableofcontents
	
\section{Introduction}
Bulk RNA sequencing (RNA-Seq) is a popular way to study gene expression mechanisms in a wide range of contexts \cite{Mortazavi2008}. The importance of RNA-Seq studies has increased significantly due to the advent of high-throughput sequencing technologies \cite{Lagstad2017}, which enable the generation of large amounts of data at a lower cost and in a shorter time than previous methods. Computational methods for processing and analyzing this data have also developed rapidly \cite{Kuehl2024, Costa-Silva2023}.

Modern high-throughput sequencing platforms, such as the Illumina HiSeq, generate millions of paired-end reads per biological sample, these reads can have between 150 and 300 base pairs in length. RNA-seq data analysis typically involves several key steps \cite{Corchete2020}. Popular steps are: I) Trimming: Removal of low-quality bases and sequencing adapters from reads; II) Filtering: Exclusion of reads with low quality or insufficient length; III) Alignment: Mapping of RNA-seq reads to a reference genome or transcriptome \cite{Patro2017SalmonExpression, trapnell2009tophat, Langmead2010}; IV) Count: Quantification of gene expression by counting reads mapped to each gene or region of interest \cite{anders2013count, Liao2019}; VI) Normalization: Adjustment of count data to correct for technical factors and ensure comparability between different samples, standard normalization methods include FPKM, TPM, RPKM, and others; VI) Differential expression analysis: Identification of genes with a significantly different expression between different conditions or groups of samples \cite{robinson2010edger, Love2014}; VII)Visualization: Presentation of results in graphical and tabular formats to facilitate comprehension \cite{Monier2019, Kucukural2019, Bunis2021}.

Several methods have been developed for each stage of this analysis \cite{Corchete2020}, and have been widely evaluated by many authors \cite{Costa-Silva2017, Liao2019, anders2015, Chowdhury2018, Overbey2021NASAData}. The most common objective of RNA-Seq data is to find differentially expressed genes (DEGs) across different conditions or groups \cite{Ren2012}. Among computational methods for identifying DEGs, some strategies consider parametric statistical distributions for expression data analysis, i.e., parametric methods. Other strategies do not take prior knowledge about the expression data into account and are called non-parametric methods. \cite{Costa-Silva2023}.

We had previously developed the method consexpression \cite{Costa-Silva2017} available in Python language. The main characteristic of consexpression is the identification of DEGs based on the wisdom of crowds theory \cite{Marbach2012}. In addition, consexpression is designed to perform four of the seven standard RNA-Seq data analysis steps mentioned above: Alignment, Count, Normalization, and Differential Expression (DE). The user only needs a sequence reads file (.fastq), a reference genome (.fasta), and annotation files (.gff) to receive a list of genes identified as differentially expressed by a mixture of five or more methods. However, due to its limitations for use with non-Python users, an alternative to DE analysis by consexpression is needed.

Here, we introduce an R package version of consexpression, named consexpressionR. Developed in the R language \cite{rproject2015}, this package isolates the DE analysis and enables R users to easily adjust parameters.


\section{Implementation}
The consexpressionR package depends on R version 4.4.0 or higher and the dependencies of seven packages that perform the differential expression analysis.

This version will only run locally from R environments where is necessary that dependencies are installed. We have provided the source code, and the latest version can be downloaded through our GitHub page \href{https://github.com/costasilvati/consexpressionR}{https://github.com/costasilvati/consexpressionR}.
The latest version can be launched using user guide instructions, which are available at \href{https://costasilvati.github.io/consexpressionR/}{https://costasilvati/github.io/consexpressionR}.

\begin{figure}[!tpb]
    \includegraphics[width=\columnwidth]{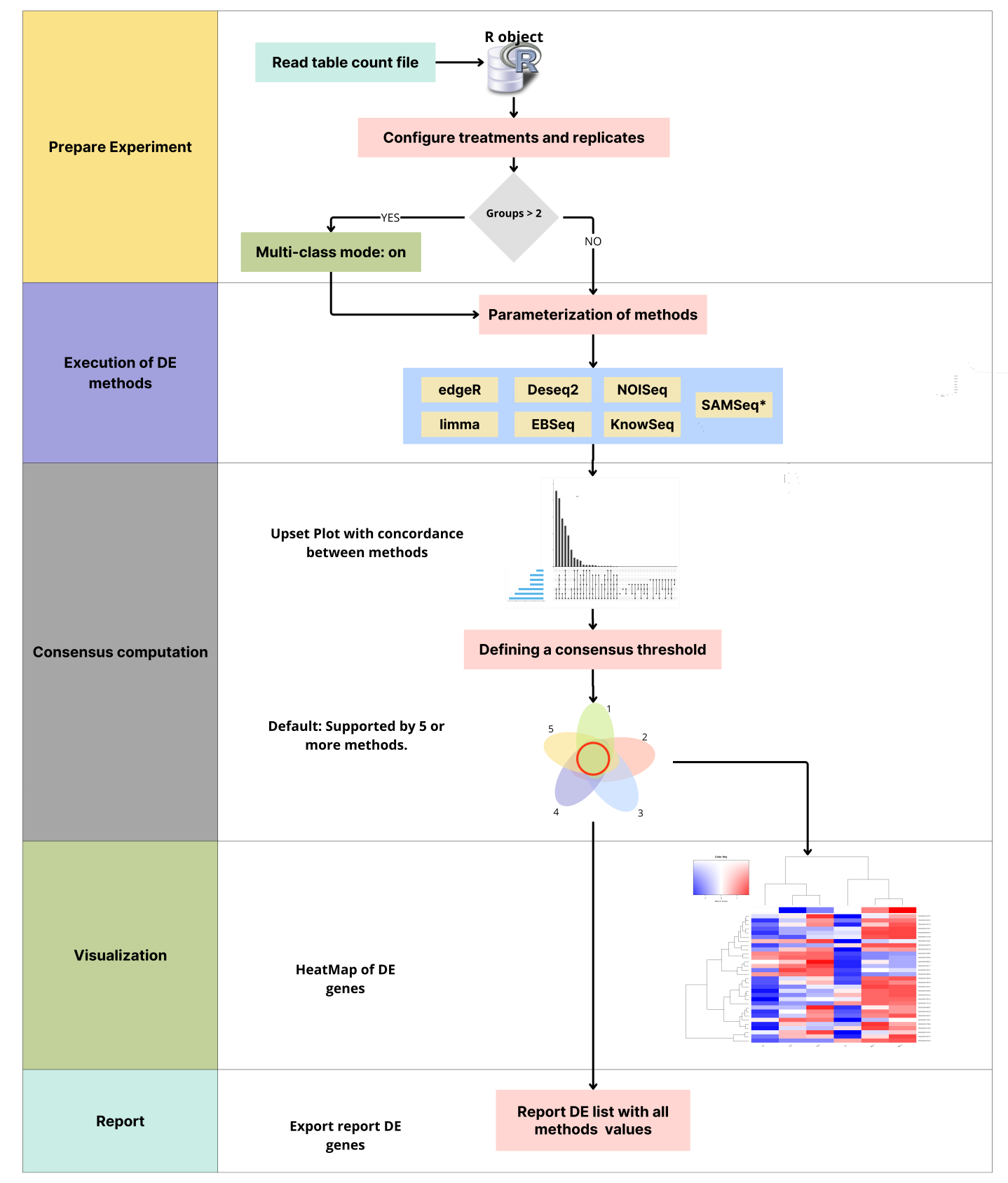}
    \caption{consexpressionR analysis workflow and main functionalities. The workflow comprises four steps, of which only visualization is optional.}\label{fig:workflow}
\end{figure}

One contribution of consexpressionR is offering a comprehensive five-stage differential expression analysis workflow. This includes preparing the experiment, reading table count files, grouping them, and replicating the assignment. The consexpressionR package makes it possible to perform 7 differential expression methods by considering their default parameters: DESeq2\cite{Love2014}, EBSeq\cite{leng2013ebseq}, edgeR\cite{robinson2010edger},  KnowSeq \cite{CastilloSecilla2021}, limma \cite{ritchie2015limma}, NOISeq\cite{tarazona2011differential}, SAMSeq\cite{li2013finding}. Consensus computation: Combine the results from different methods to obtain more robust and reliable results of gene expression changes. Visualization: Generate informative graphical summaries of your results for straightforward interpretation.\\
Report generation: Get a report summarizing all the tools used.
For a streamlined workflow (Figure: \ref{fig:workflow}), a "simple DE procedure" is available. This removes the step of visualization. Other steps are considered mandatory because they provide the essential information needed for the DE consensus analysis.

\section{Materials and Methods} \label{sec:meth}

Among the methodologies identified in the study \cite{Costa-Silva2023}, the consexpression \cite{Costa-Silva2017} stands out. This is a hybrid methodology that utilizes both parametric and non-parametric methods for the definition of DEGs, proving to be pioneering. Therefore, this study implemented the updated version of consexpression, named consexpressionR \cite{costa-silva2024gitConsexpressionR},
which presents source code, documentation for use, and all the files available for the complete replication of this study in an open way.

\subsection*{Implementation}
For the implementation of the R package consexpressionR, the characteristics of consensus use in the definition of DEGs, and the main contribution of the initial version (implemented in the Python programming language \cite{Python2024}), were taken into account. However, some updates and needs presented by users in the repository of the initial version \cite{costa-silva2017gitConsexpression} were considered in this new version. These are:

\begin{enumerate}
    \item Updates: new versions of R packages and other extensions used can cause execution problems by changing the form of commands and parameters, for example;
    \item Need for partial analysis: the initial version of consexpression performs all the steps of expression analysis (mapping, counting, and identifying DEGs). In the new version, these steps have been implemented independently, and only the identification of DEGs has been made available;
    \item Analyses for organisms without a reference genome: consexpression requires the FASTA file with the reference genome and the GFF file with genome annotation for its execution, which prevents users with \textit{de novo} data from performing their analysis. The new version provides differential expression analysis without the need for a reference genome;
    \item Usability: consexpression does not have a graphical user interface (GUI), which can be a limiting factor for the use of the tool,  which was made available in the new version of the package.
\end{enumerate}

In this context, the implementation of consexpressionR follows the R package standard, as presented in the book \citealp{Wickman2015Livros}. For this purpose, the package supporting the development of R packages, \emph{devtools} \cite{devtools2022}, was used to assist in creating the structure. The documentation was developed using the \emph{roxygen2} and \emph{devtools} package \cite{roxygen2024}. The graphical interface was developed using the R Shiny package \cite{Winston2024shiny}.

\begin{table}[!htp]
\centering
\caption{DEGs methods adopted in R package consexpressionR, order by year of publication. It adopted the methodology proposed in review \cite{Costa-Silva2023}.}
\label{tab:consexpressioR-packageDEG}
\begin{tabular}{c|c|c} \hline
Method & Reference & Classification \\ \hline 
edger  & \cite{robinson2010edger} &  Parametric \\ \hline
NOISeq & \cite{tarazona2011differential} &  Non parametric \\ \hline
EBSeq  & \cite{leng2013ebseq} & Parametric \\ \hline
SAMSeq & \cite{li2013finding} &  Non parametric \\ \hline
DESeq2 & \cite{Love2014} &  Parametric \\ \hline
limma  & \cite{ritchie2015limma} &  Parametric \\ \hline
KnowSeq & \cite{CastilloSecilla2021} & Parametric \\ \hline
\end{tabular}
\end{table}

To address the gaps identified in the consexpression, the analysis of consexpressionR starts from the integer values of the count table, as presented in Figure \ref{fig:workflow}. The analysis of DEGs is performed with seven methods implemented in R package format, namely: edgeR, DESeq2, baySeq, EBSeq, NOISeq, limma, and knowSeq, as presented in Table \ref{tab:consexpressioR-packageDEG}. For the SAMSeq method, only execution with count data is allowed; this method is not executed for quantification data. 
The KnowSeq method is executed only for data with a reference genome and valid gene names in public annotation data, such as ENSEMBL \cite{Mudunuri2009}.

\subsection*{Datasets}

Two datasets were adopted in order to evaluate the differential expression in this study.
Dataset (A) is available in the Short Read Archive (SRA) by the National Center for Biotechnology Information (NCBI) and can be accessed using the code SRA010153. This dataset includes two conditions: cerebral tissue (Brain) and a mix of human cells (UHR), with each condition having seven replicates \cite{bullard2010evaluation}. The reads of this dataset were mapped to the human genome, version 19 (GRCh37.p13) using the TopHat tool \cite{trapnell2009tophat}. The genome and annotation file used are available in GENCODE page project \cite{harrow2012gencode}.  To performance metrics were used as \textit{gold standard} experiment, the qPCR constructed with the same samples described. This is the same dataset used in the version of consexpression \cite{Costa-Silva2017}.

Dataset (B) is available in the Gene Expression Omnibus (GEO) repository of NCBI \cite{Barrett2011NCBIOn} and can accessed using the code GSE95077. This dataset provides the count table from an experiment conducted using the Illumina HiSeq 2500 sequencer. The experiment involved two conditions: multiple myeloma cells with two treatments (BM and JJ), representing the application of amiloride at different dosages, and a control (DMSO). Each condition contains six samples, resulting in a count table with 18 columns and 107 rows. Performance metrics were obtained using qPCR data from the same study \cite{Corchete2020}. 31 genes were considered eligible for expression analysis with qPCR. These genes, $\frac{1}{3}$, were randomly chosen, and $\frac{2}{3}$ were selected because they showed significant expression variation, such as being up or downregulated. For the study initially applied, dataset B was ranked. However, in this study, we consider that genes classified by \citealp{Corchete2020} as upregulated and downregulated by qPCR are differentially expressed, regardless of their ranking position.

\section{Results}
In order to evaluate the proposed method to identify DEGs, an experiment was performed using data sets (A) and (B), detailed information about data sets is presented in Section \ref{sec:meth}. Various assertiveness metrics were considered, as shown in Table \ref{tab:consesnus_resultA}.

The evaluation of individual methodologies for identifying DEGs clearly indicates that the results are strongly influenced by the adopted statistical model. Some methods perform better with more samples, while others show variations in the results influenced by different study characteristics \cite{Costa-Silva2017}.

\begin{table}[!htp]
\centering
\caption{Performance evaluation of consensus with dataset (A).}
\label{tab:consesnus_resultA}
\begin{tabular}{c|c|c|c|c|c}
\hline
\textbf{Consensus \#}& \textbf{TP} & \textbf{FP} & \textbf{Recall} & \textbf{Specificity} & \textbf{Precision} \\ \hline
\textbf{1} & 371    & 354   & 0,93  & 0,36  & 0,51  \\ \hline
\textbf{2} & 359    & 226   & 0,90  & 0,59  & 0,61  \\ \hline
\textbf{3} & 342    & 80    & 0,86  & 0,85  & 0,81  \\ \hline
\textbf{4} & 333    & 38    & 0,83  & 0,93  & 0,89  \\ \hline
\textbf{5} & 310    & 20    & 0,78  & 0,96  & 0,94  \\ \hline
\textbf{6} & 263    & 8     & 0,66  & 0,98  & 0,97  \\ \hline
\textbf{7} & 87     & 6     & 0,22  & 0,99  & 0,93  \\ \hline
\end{tabular}
\end{table}

We applied seven differential gene expression analysis techniques (presented in Table \ref{tab:consexpressioR-packageDEG}) to the data used in the initial consexpression tests \cite{Costa-Silva2017}, here referred to as dataset (A) and to data from the study by \citealp{Corchete2020}, here referred to as dataset (B). We evaluated the consensus of these results to determine whether the updated theory maintains the previously recorded performance.

\begin{table}[!htp]
\centering
\caption{Performance evaluation of consensus with dataset (B).}
\label{tab:consesnus_resultB}
\begin{tabular}{c|c|c|c|c|c}
\hline
  \textbf{Consensus \#} & \textbf{TP} & \textbf{FP} & \textbf{Recall} & \textbf{Specificity} & \textbf{Precision} \\ \hline
1 & 13 & 7  & 0,62   & 0,30        & 0,65      \\ \hline
2 & 11 & 4  & 0,52   & 0,60        & 0,73      \\ \hline
3 & 7  & 1  & 0,33   & 0,90        & 0,87      \\ \hline
4 & 4  & 0  & 0,19   & 1,00        & 1,00      \\ \hline
5 & 4  & 0  & 0,19   & 1,00        & 1,00      \\ \hline
6 & 3  & 0  & 0,14   & 1,00        & 1,00      \\ \hline
7 & 0  & 0  & 0,00   & 1,00        & NA        \\ \hline
\end{tabular}
\end{table}

Tables \ref{tab:consesnus_resultA} and \ref{tab:consesnus_resultB} present the performance measures of DEG analysis using the consensus on the range of one to seven methodologies. Dataset (A) follows a normal distribution for RNA-Seq expression experiments, where most data do not show significant expression variations and a smaller group shows evident variation. 

Consensus analysis of dataset (A) reveals that consensus among four or five methods yields results comparable to the best individual method performance while improving precision.  However, using a consensus of more than five methods can lead to high precision but low recall. This approach may be helpful for identifying specific variations but at the cost of potentially missing relevant discoveries.

\begin{figure}[!htb]
\centerline{\includegraphics[width=0.5\textwidth]{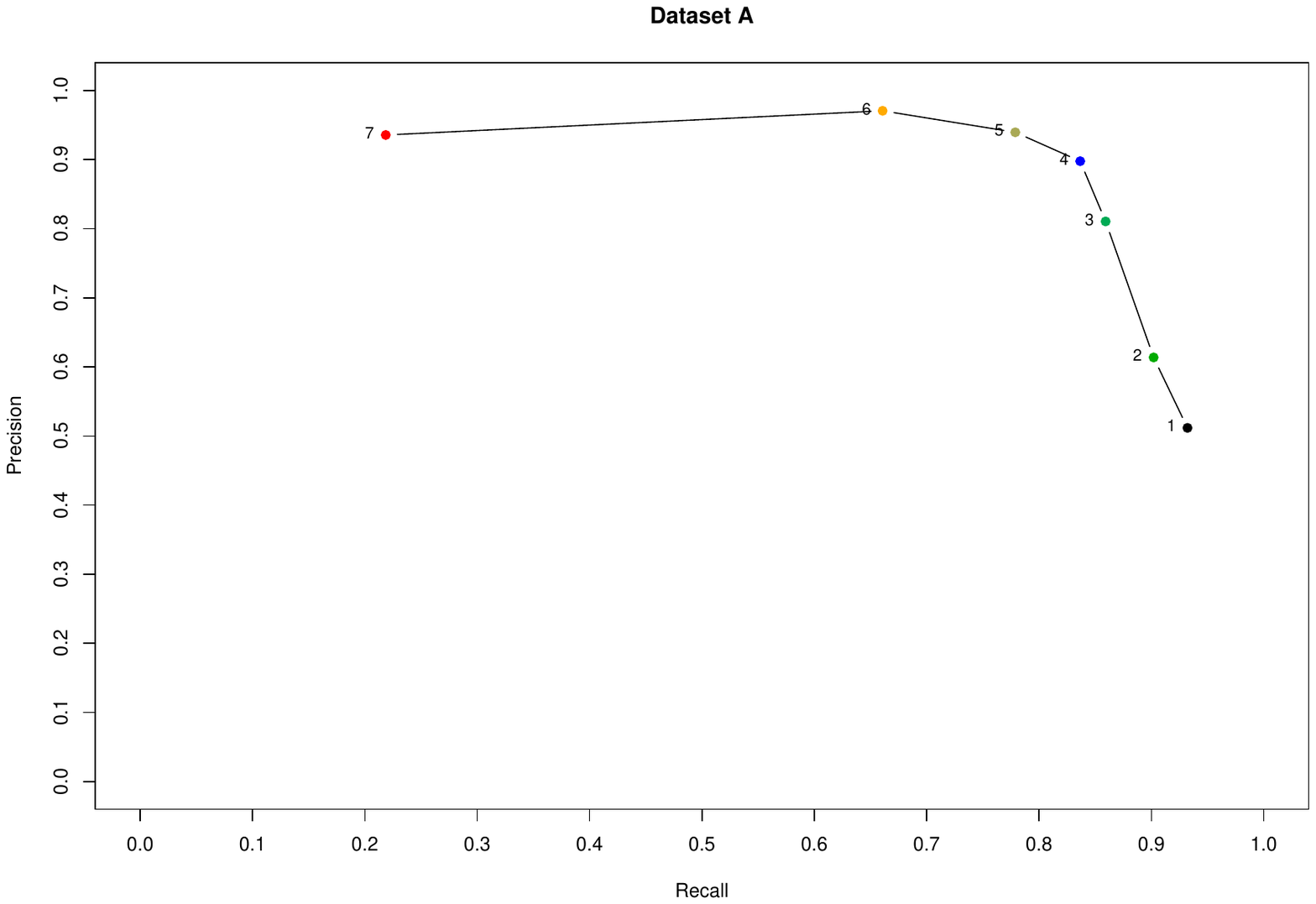}}
\caption{Precision X Recall curve by dataset (A) by considering the consensus of methods.}\label{fig:01}
\end{figure}

\begin{figure}[!htb]
\centerline{\includegraphics[width=0.5\textwidth]{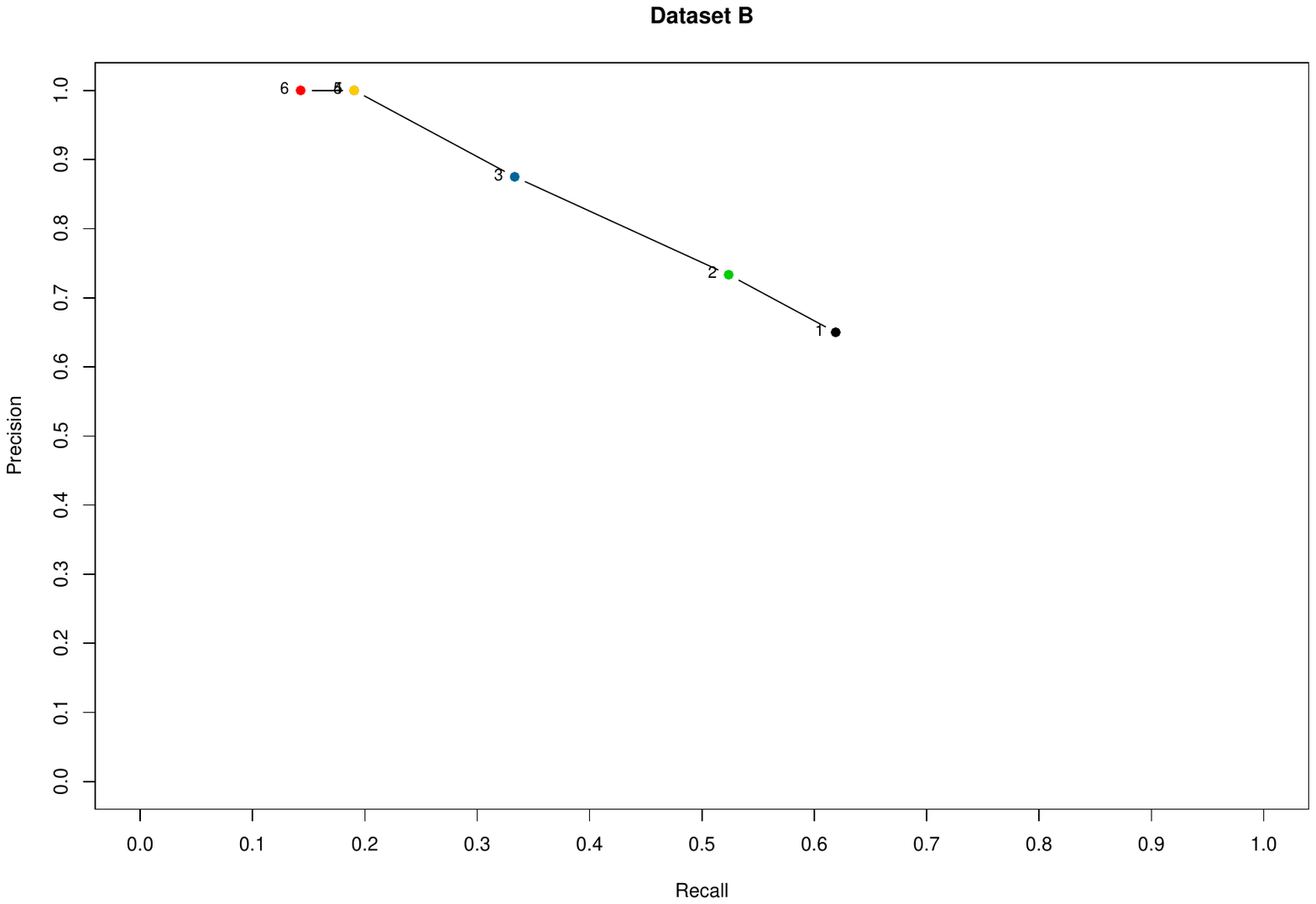}}
\caption{Precision X Recall curve by dataset (B) by considering the consensus of methods.}\label{fig:02}
\end{figure}

The precision of the consensus methods improves with each added methodology, as shown in figures \ref{fig:01} and \ref{fig:02}. However, when the specificity is increased, the result becomes highly restrictive, so that in a data set with few genes, such as dataset (B), the FP ratio reaches 0 when we select only the genes indicated as DE by at least four methods.

Dataset (B) analysis shows that while EBSeq and NOISeq achieve high recall, they also have the three lowest precision rates, indicating a high rate of false positives despite identifying numerous genes. These results are consistent with those observed in dataset (A). Among the seven methods evaluated, edgeR, limma, DESeq2, and SAMSeq achieve the highest accuracy, consistent with previous studies \cite{Corchete2020SystematicAnalysis, Costa-Silva2017}. This suggests that nearly all genes identified by these methods are true positives, given the default parameterization used by consexpressionR. However, the limited number of samples leads to lower recall.

The results indicate that consensus analysis improves the identification of differentially expressed genes, yielding more robust and reliable results. This approach facilitates the rapid generation of gene lists based on the agreement of $N$ user-specified methods (ranging from one to seven), irrespective of the specific methods employed. Notably, the final gene list may comprise genes identified by $N$ methods, even if none of those methods exhibits optimal individual performance. Thus, the parameter $N$ can be adjusted in order to produce results with higher precision (avoiding false positives), more permissive results with greater recovery (recall), or even a balance between precision and recall.

\section{Conclusion}

In this work, we present the R package, consexpressionR, which leverages the consensus of multiple differential expression analysis methods to identify DEGs. We also analyzed the individual results of the main methods across two RNA-Seq benchmark datasets. The experiments indicate that combining multiple methods can enhance the specificity of DEG identification; Thus, consexpressionR allows parametrization to enable more sensitive analyses as well.

As noted previously \cite{Costa-Silva2017}, we detected a decrease in Recall as the number of methods in consensus; in datasets (A) and (B), it is possible to see the same behavior. This range follows an increase in precision rate. Thus, as a result, using more consensus methods can indicate DEGs more precisely and restrictively. On the other hand, discovering new DEGs requires being more permissive.
In individual methods analysis, it is possible to identify that non-parametric methods are more affected by the amount of data. NOISeq and KnowSeq have a low Recall rate. 

A graphical user interface (GUI) was developed for the package, enabling local execution of analysis through a web browser. The package is currently available on GitHub.  We anticipate its inclusion in the Bioconductor repository following the completion of the submission process.

Based on the results of the experiments, we conclude that RNA-Seq-based differential expression methodologies have achieved a high level of maturity. Therefore, current contributions focus on improving ease of use and enhancing the robustness of existing methods. Future research directions include exploring consensus-based analysis of single-cell RNA-Seq (scRNA-Seq) data.

In summary, consexpressionR provides a reliable R package for Bulk RNA-Seq expression analysis, delivering highly specific results. Additionally, it offers the option to identify new potential expression profiles when using a lower consensus threshold, such as the indication of one or two methods.

\section{Funding}
This study was financed by the Fundação Araucária (Grant number 035/2019, 138/2021 and NAPI - Bioinformática) and CNPq (Grant number 440412/2022-6 and 408312/2023-8).

%
%
\bibliography{ref}

\begin{thebibliography}{39}
\providecommand{\natexlab}[1]{#1}
\providecommand{\url}[1]{\texttt{#1}}
\expandafter\ifx\csname urlstyle\endcsname\relax
  \providecommand{\doi}[1]{doi: #1}\else
  \providecommand{\doi}{doi: \begingroup \urlstyle{rm}\Url}\fi

\bibitem[Anders et~al.(2013)Anders, McCarthy, Chen, Okoniewski, Smyth, Huber, and Robinson]{anders2013count}
Simon Anders, Davis~J McCarthy, Yunshun Chen, Michal Okoniewski, Gordon~K Smyth, Wolfgang Huber, and Mark~D Robinson.
\newblock Count-based differential expression analysis of rna sequencing data using r and bioconductor.
\newblock \emph{Nature protocols}, 8:\penalty0 1765--1786, 2013.

\bibitem[Anders et~al.(2015)Anders, Pyl, and Huber]{anders2015}
Simon Anders, Paul~Theodor Pyl, and Wolfgang Huber.
\newblock Htseq-a python framework to work with high-throughput sequencing data.
\newblock \emph{Bioinformatics}, 31:\penalty0 166--169, 1 2015.
\newblock ISSN 14602059.
\newblock \doi{10.1093/bioinformatics/btu638}.

\bibitem[Barrett et~al.(2011)Barrett, Troup, Wilhite, Ledoux, Evangelista, Kim, Tomashevsky, Marshall, Phillippy, Sherman, Muertter, Holko, Ayanbule, Yefanov, and Soboleva]{Barrett2011NCBIOn}
Tanya Barrett, Dennis~B. Troup, Stephen~E. Wilhite, Pierre Ledoux, Carlos Evangelista, Irene~F. Kim, Maxim Tomashevsky, Kimberly~A. Marshall, Katherine~H. Phillippy, Patti~M. Sherman, Rolf~N. Muertter, Michelle Holko, Oluwabukunmi Ayanbule, Andrey Yefanov, and Alexandra Soboleva.
\newblock Ncbi geo: archive for functional genomics data sets—10 years on.
\newblock \emph{Nucleic Acids Research}, 39\penalty0 (1):\penalty0 D1005--D1010, 1 2011.
\newblock ISSN 0305-1048.
\newblock \doi{10.1093/NAR/GKQ1184}.
\newblock URL \url{https://dx.doi.org/10.1093/nar/gkq1184}.

\bibitem[Bullard et~al.(2010)Bullard, Purdom, Hansen, and Dudoit]{bullard2010evaluation}
James~H Bullard, Elizabeth Purdom, Kasper~D Hansen, and Sandrine Dudoit.
\newblock Evaluation of statistical methods for normalization and differential expression in mrna-seq experiments.
\newblock \emph{BMC Bioinformatics}, 11\penalty0 (1):\penalty0 94, 12 2010.
\newblock ISSN 1471-2105.
\newblock \doi{10.1186/1471-2105-11-94}.
\newblock URL \url{https://bmcbioinformatics.biomedcentral.com/articles/10.1186/1471-2105-11-94}.

\bibitem[Bunis et~al.(2021)Bunis, Andrews, Fragiadakis, Burt, and Sirota]{Bunis2021}
Daniel~G Bunis, Jared Andrews, Gabriela~K Fragiadakis, Trevor~D Burt, and Marina Sirota.
\newblock dittoseq: universal user-friendly single-cell and bulk rna sequencing visualization toolkit.
\newblock \emph{Bioinformatics}, 36:\penalty0 5535--5536, 4 2021.
\newblock ISSN 1367-4803.
\newblock \doi{10.1093/bioinformatics/btaa1011}.
\newblock URL \url{https://academic.oup.com/bioinformatics/article/36/22-23/5535/6031909}.

\bibitem[Castillo-Secilla et~al.(2021)Castillo-Secilla, G{\'{a}}lvez, Carrillo-Perez, Verona-Almeida, Redondo-S{\'{a}}nchez, Ortuno, Herrera, and Rojas]{CastilloSecilla2021}
Daniel Castillo-Secilla, Juan~Manuel G{\'{a}}lvez, Francisco Carrillo-Perez, Marta Verona-Almeida, Daniel Redondo-S{\'{a}}nchez, Francisco~Manuel Ortuno, Luis~Javier Herrera, and Ignacio Rojas.
\newblock Knowseq r-bioc package: The automatic smart gene expression tool for retrieving relevant biological knowledge.
\newblock \emph{Computers in Biology and Medicine}, 133:\penalty0 104387, 6 2021.
\newblock ISSN 0010-4825.
\newblock \doi{10.1016/J.COMPBIOMED.2021.104387}.

\bibitem[Chang et~al.(2024)Chang, Cheng, Allaire, Sievert, Schloerke, Xie, Allen, McPherson, Dipert, and Borges]{Winston2024shiny}
Winston Chang, Joe Cheng, JJ~Allaire, Carson Sievert, Barret Schloerke, Yihui Xie, Jeff Allen, Jonathan McPherson, Alan Dipert, and Barbara Borges.
\newblock \emph{shiny: Web Application Framework for R}, 2024.
\newblock URL \url{https://shiny.posit.co/}.
\newblock R package version 1.9.0.9000, https://github.com/rstudio/shiny.

\bibitem[Chowdhury et~al.(2020)Chowdhury, Bhattacharyya, and Kalita]{Chowdhury2018}
Hussain~Ahmed Chowdhury, Dhruba~Kumar Bhattacharyya, and Jugal~Kumar Kalita.
\newblock Differential expression analysis of rna-seq reads: Overview, taxonomy, and tools.
\newblock \emph{IEEE/ACM Transactions on Computational Biology and Bioinformatics}, 17\penalty0 (2):\penalty0 566--586, 3 2020.
\newblock ISSN 15579964.
\newblock \doi{10.1109/TCBB.2018.2873010}.

\bibitem[Corchete et~al.(2020{\natexlab{a}})Corchete, Rojas, Alonso-L{\'{o}}pez, De~Las~Rivas, Guti{\'{e}}rrez, and Burguillo]{Corchete2020SystematicAnalysis}
Luis~A. Corchete, Elizabeta~A. Rojas, Diego Alonso-L{\'{o}}pez, Javier De~Las~Rivas, Norma~C. Guti{\'{e}}rrez, and Francisco~J. Burguillo.
\newblock Systematic comparison and assessment of rna-seq procedures for gene expression quantitative analysis.
\newblock \emph{Scientific Reports}, 10\penalty0 (1), 12 2020{\natexlab{a}}.
\newblock ISSN 20452322.
\newblock \doi{10.1038/s41598-020-76881-x}.

\bibitem[Corchete et~al.(2020{\natexlab{b}})Corchete, Rojas, Alonso-López, Rivas, Gutiérrez, and Burguillo]{Corchete2020}
Luis~A. Corchete, Elizabeta~A. Rojas, Diego Alonso-López, Javier De~Las Rivas, Norma~C. Gutiérrez, and Francisco~J. Burguillo.
\newblock Systematic comparison and assessment of rna-seq procedures for gene expression quantitative analysis.
\newblock \emph{Nature Scientific Reports}, 10:\penalty0 19737, 11 2020{\natexlab{b}}.
\newblock ISSN 2045-2322.
\newblock \doi{10.1038/s41598-020-76881-x}.
\newblock URL \url{https://www.nature.com/articles/s41598-020-76881-x}.

\bibitem[Costa-Silva et~al.(2017{\natexlab{a}})Costa-Silva, Domingues, and Lopes]{Costa-Silva2017}
Juliana Costa-Silva, Douglas Domingues, and Fabricio~Martins Lopes.
\newblock Rna-seq differential expression analysis: An extended review and a software tool.
\newblock \emph{PLOS ONE}, 12\penalty0 (12):\penalty0 e0190152, 12 2017{\natexlab{a}}.
\newblock ISSN 1932-6203.
\newblock \doi{10.1371/journal.pone.0190152}.
\newblock URL \url{https://dx.plos.org/10.1371/journal.pone.0190152}.

\bibitem[Costa-Silva et~al.(2017{\natexlab{b}})Costa-Silva, Menotti, and Lopes]{costa-silva2017gitConsexpression}
Juliana Costa-Silva, Douglas~S. Menotti, Domingues, and Fabrício~M. Lopes.
\newblock costasilvati/consexpression:differential gene expression software with consesnus, 2017{\natexlab{b}}.
\newblock URL \url{https://github.com/costasilvati/consexpression}.

\bibitem[Costa-Silva et~al.(2023)Costa-Silva, Domingues, Menotti, Hungria, and Lopes]{Costa-Silva2023}
Juliana Costa-Silva, Douglas~S. Domingues, David Menotti, Mariangela Hungria, and Fabrício~Martins Lopes.
\newblock Temporal progress of gene expression analysis with rna-seq data: A review on the relationship between computational methods.
\newblock \emph{Computational and Structural Biotechnology Journal}, 21:\penalty0 86--98, 1 2023.
\newblock ISSN 2001-0370.
\newblock \doi{10.1016/J.CSBJ.2022.11.051}.

\bibitem[Costa-Silva et~al.(2024)Costa-Silva, Menotti, and Lopes]{costa-silva2024gitConsexpressionR}
Juliana Costa-Silva, David~Gomes Menotti, and Fabricio~M. Lopes.
\newblock costasilvati/consexpressionr: Version 2.0 of differential gene expression software, 2024.
\newblock URL \url{https://github.com/costasilvati/consexpressionR}.

\bibitem[Foundation(2024)]{Python2024}
Python~Software Foundation.
\newblock \emph{Python.org}, 2024.
\newblock URL \url{https://www.python.org/}.

\bibitem[Harrow et~al.(2012)Harrow, Frankish, Gonzalez, Tapanari, Diekhans, Searle, and {others}]{harrow2012gencode}
Jennifer Harrow, Adam Frankish, Jose~M Gonzalez, Electra Tapanari, Mark Diekhans, Stephen Searle, and {others}.
\newblock Gencode: the reference human genome annotation for the encode project.
\newblock \emph{Genome research}, 22\penalty0 (9):\penalty0 1760--1774, 2012.

\bibitem[Hornik(2014)]{rproject2015}
Kurt Hornik.
\newblock R: The r project for statistical computing, 2014.
\newblock URL \url{http://www.r-project.org}.

\bibitem[Kucukural et~al.(2019)Kucukural, Yukselen, Ozata, Moore, and Garber]{Kucukural2019}
Alper Kucukural, Onur Yukselen, Deniz~M. Ozata, Melissa~J. Moore, and Manuel Garber.
\newblock Debrowser: Interactive differential expression analysis and visualization tool for count data 06 biological sciences 0604 genetics 08 information and computing sciences 0806 information systems.
\newblock \emph{BMC Genomics}, 20\penalty0 (1):\penalty0 6, 1 2019.
\newblock ISSN 14712164.
\newblock \doi{10.1186/s12864-018-5362-x}.
\newblock URL \url{https://bmcgenomics.biomedcentral.com/articles/10.1186/s12864-018-5362-x}.

\bibitem[Kuehl et~al.(2024)Kuehl, Wong, Wanner, Bonn, and Puelles]{Kuehl2024}
Malte Kuehl, Milagros~N. Wong, Nicola Wanner, Stefan Bonn, and Victor~G. Puelles.
\newblock Gene count estimation with pytximport enables reproducible analysis of bulk rna sequencing data in python.
\newblock \emph{Bioinformatics}, 40, 11 2024.
\newblock ISSN 13674811.
\newblock \doi{10.1093/BIOINFORMATICS/BTAE700}.
\newblock URL \url{https://dx.doi.org/10.1093/bioinformatics/btae700}.

\bibitem[Langmead et~al.(2010)Langmead, Hansen, and Leek]{Langmead2010}
Ben Langmead, Kasper~D. Hansen, and Jeffrey~T. Leek.
\newblock Cloud-scale rna-sequencing differential expression analysis with myrna.
\newblock \emph{Genome biology}, 11\penalty0 (8):\penalty0 1--11, 8 2010.
\newblock ISSN 14656914.
\newblock \doi{10.1186/gb-2010-11-8-r83}.
\newblock URL \url{https://link.springer.com/articles/10.1186/gb-2010-11-8-r83 https://link.springer.com/article/10.1186/gb-2010-11-8-r83}.

\bibitem[Leng et~al.(2013)Leng, Dawson, Thomson, Ruotti, Rissman, Smits, Haag, Gould, Stewart, and Kendziorski]{leng2013ebseq}
Ning Leng, John~A. Dawson, James~A. Thomson, Victor Ruotti, Anna~I. Rissman, Bart M.~G. Smits, Jill~D. Haag, Michael~N. Gould, Ron~M. Stewart, and Christina Kendziorski.
\newblock Ebseq: an empirical bayes hierarchical model for inference in rna-seq experiments.
\newblock \emph{Bioinformatics}, 29\penalty0 (8):\penalty0 1035--1043, 4 2013.
\newblock ISSN 1460-2059.
\newblock \doi{10.1093/bioinformatics/btt087}.
\newblock URL \url{https://academic.oup.com/bioinformatics/article-lookup/doi/10.1093/bioinformatics/btt087}.

\bibitem[Li and Tibshirani(2013)]{li2013finding}
Jun Li and Robert Tibshirani.
\newblock Finding consistent patterns: A nonparametric approach for identifying differential expression in rna-seq data.
\newblock \emph{Statistical Methods in Medical Research}, 22\penalty0 (5):\penalty0 519--536, 10 2013.
\newblock ISSN 0962-2802.
\newblock \doi{10.1177/0962280211428386}.
\newblock URL \url{http://journals.sagepub.com/doi/10.1177/0962280211428386}.

\bibitem[Liao et~al.(2019)Liao, Smyth, and Shi]{Liao2019}
Yang Liao, Gordon~K Smyth, and Wei Shi.
\newblock The r package rsubread is easier, faster, cheaper and better for alignment and quantification of rna sequencing reads.
\newblock \emph{Nucleic Acids Research}, 47\penalty0 (8):\penalty0 e47--e47, 5 2019.
\newblock ISSN 0305-1048.
\newblock \doi{10.1093/nar/gkz114}.
\newblock URL \url{https://academic.oup.com/nar/article/47/8/e47/5345150}.

\bibitem[Love et~al.(2014)Love, Huber, and Anders]{Love2014}
Michael~I Love, Wolfgang Huber, and Simon Anders.
\newblock Moderated estimation of fold change and dispersion for rna-seq data with deseq2.
\newblock \emph{Genome Biology}, 15\penalty0 (12):\penalty0 550, 12 2014.
\newblock ISSN 1474-760X.
\newblock \doi{10.1186/s13059-014-0550-8}.
\newblock URL \url{http://genomebiology.biomedcentral.com/articles/10.1186/s13059-014-0550-8}.

\bibitem[Lågstad et~al.(2017)Lågstad, Zhao, Hoff, Johannessen, Lingjærde, and Skotheim]{Lagstad2017}
Stian Lågstad, Sen Zhao, Andreas~M. Hoff, Bjarne Johannessen, Ole~Christian Lingjærde, and Rolf~I. Skotheim.
\newblock chimeraviz: a tool for visualizing chimeric rna.
\newblock \emph{Bioinformatics}, 33:\penalty0 2954--2956, 9 2017.
\newblock ISSN 1367-4803.
\newblock \doi{10.1093/BIOINFORMATICS/BTX329}.
\newblock URL \url{https://dx.doi.org/10.1093/bioinformatics/btx329}.

\bibitem[Marbach et~al.(2012)Marbach, Costello, K{\"{u}}ffner, Vega, Prill, Camacho, Allison, {DREAM5 Consortium}, Kellis, Collins, and Stolovitzky]{Marbach2012}
Daniel Marbach, James~C Costello, Robert K{\"{u}}ffner, Nicole~M Vega, Robert~J Prill, Diogo~M Camacho, Kyle~R Allison, {DREAM5 Consortium}, Manolis Kellis, James~J Collins, and Gustavo Stolovitzky.
\newblock Wisdom of crowds for robust gene network inference.
\newblock \emph{Nature methods}, 9\penalty0 (8):\penalty0 796--804, 7 2012.
\newblock ISSN 1548-7105.
\newblock \doi{10.1038/nmeth.2016}.
\newblock URL \url{http://www.ncbi.nlm.nih.gov/pubmed/22796662}.

\bibitem[Monier et~al.(2019)Monier, McDermaid, Wang, Zhao, Miller, Fennell, and Ma]{Monier2019}
Brandon Monier, Adam McDermaid, Cankun Wang, Jing Zhao, Allison Miller, Anne Fennell, and Qin Ma.
\newblock Iris-eda: An integrated rna-seq interpretation system for gene expression data analysis.
\newblock \emph{PLoS Computational Biology}, 15, 2 2019.
\newblock ISSN 15537358.
\newblock \doi{10.1371/journal.pcbi.1006792}.

\bibitem[Mortazavi et~al.(2008)Mortazavi, Williams, McCue, Schaeffer, and Wold]{Mortazavi2008}
Ali Mortazavi, Brian~A. Williams, Kenneth McCue, Lorian Schaeffer, and Barbara Wold.
\newblock Mapping and quantifying mammalian transcriptomes by rna-seq.
\newblock \emph{Nature Methods}, 5\penalty0 (7):\penalty0 621--628, 7 2008.
\newblock ISSN 15487091.
\newblock \doi{10.1038/nmeth.1226}.

\bibitem[Mudunuri et~al.(2009)Mudunuri, Che, Yi, and Stephens]{Mudunuri2009}
Uma Mudunuri, Anney Che, Ming Yi, and Robert~M. Stephens.
\newblock {bioDBnet: the biological database network}.
\newblock \emph{Bioinformatics}, 25\penalty0 (4):\penalty0 555--556, 01 2009.
\newblock ISSN 1367-4803.
\newblock \doi{10.1093/bioinformatics/btn654}.
\newblock URL \url{https://doi.org/10.1093/bioinformatics/btn654}.

\bibitem[Overbey et~al.(2021)Overbey, Saravia-Butler, Zhang, Rathi, Fogle, da~Silveira, Barker, Bass, Beheshti, Berrios, Blaber, Cekanaviciute, Costa, Davin, Fisch, Gebre, Geniza, Gilbert, Gilroy, Hardiman, Herranz, Kidane, Kruse, Lee, Liefeld, Lewis, McDonald, Meller, Mishra, Perera, Ray, Reinsch, Rosenthal, Strong, Szewczyk, Tahimic, Taylor, Vandenbrink, Villacampa, Weging, Wolverton, Wyatt, Zea, Costes, and Galazka]{Overbey2021NASAData}
Eliah~G. Overbey, Amanda~M. Saravia-Butler, Zhe Zhang, Komal~S. Rathi, Homer Fogle, Willian~A. da~Silveira, Richard~J. Barker, Joseph~J. Bass, Afshin Beheshti, Daniel~C. Berrios, Elizabeth~A. Blaber, Egle Cekanaviciute, Helio~A. Costa, Laurence~B. Davin, Kathleen~M. Fisch, Samrawit~G. Gebre, Matthew Geniza, Rachel Gilbert, Simon Gilroy, Gary Hardiman, Raúl Herranz, Yared~H. Kidane, Colin~P.S. Kruse, Michael~D. Lee, Ted Liefeld, Norman~G. Lewis, J.~Tyson McDonald, Robert Meller, Tejaswini Mishra, Imara~Y. Perera, Shayoni Ray, Sigrid~S. Reinsch, Sara~Brin Rosenthal, Michael Strong, Nathaniel~J. Szewczyk, Candice~G.T. Tahimic, Deanne~M. Taylor, Joshua~P. Vandenbrink, Alicia Villacampa, Silvio Weging, Chris Wolverton, Sarah~E. Wyatt, Luis Zea, Sylvain~V. Costes, and Jonathan~M. Galazka.
\newblock Nasa genelab rna-seq consensus pipeline: standardized processing of short-read rna-seq data.
\newblock \emph{iScience}, 24\penalty0 (4):\penalty0 102361, 4 2021.
\newblock ISSN 25890042.
\newblock \doi{10.1016/J.ISCI.2021.102361/ATTACHMENT/A6434F11-57A4-4B99-A3B6-718D5BDB0F99/MMC8.ZIP}.
\newblock URL \url{http://www.cell.com/article/S2589004221003291/fulltext}.

\bibitem[Patro et~al.(2017)Patro, Duggal, Love, Irizarry, and Kingsford]{Patro2017SalmonExpression}
Rob Patro, Geet Duggal, Michael~I Love, Rafael~A Irizarry, and Carl Kingsford.
\newblock Salmon provides fast and bias-aware quantification of transcript expression.
\newblock \emph{Nature Methods}, 14\penalty0 (4):\penalty0 417--419, 4 2017.
\newblock ISSN 1548-7091.
\newblock \doi{10.1038/nmeth.4197}.
\newblock URL \url{http://www.nature.com/articles/nmeth.4197}.

\bibitem[Ren et~al.(2012)Ren, Peng, Mao, Yu, Yin, Gao, Cui, Zhang, Yi, Xu, Chen, Wang, Guo, Lu, Yang, Wei, Tian, Guan, Tang, Xu, Wang, Gao, Tian, Wang, Yang, Wang, and Sun]{Ren2012}
Shancheng Ren, Zhiyu Peng, Jian-Hua Mao, Yongwei Yu, Changjun Yin, Xin Gao, Zilian Cui, Jibin Zhang, Kang Yi, Weidong Xu, Chao Chen, Fubo Wang, Xinwu Guo, Ji~Lu, Jun Yang, Min Wei, Zhijian Tian, Yinghui Guan, Liang Tang, Chuanliang Xu, Linhui Wang, Xu~Gao, Wei Tian, Jian Wang, Huanming Yang, Jun Wang, and Yinghao Sun.
\newblock Rna-seq analysis of prostate cancer in the chinese population identifies recurrent gene fusions, cancer-associated long noncoding rnas and aberrant alternative splicings.
\newblock \emph{Cell Research}, 22\penalty0 (5):\penalty0 806--821, 5 2012.
\newblock ISSN 1001-0602.
\newblock \doi{10.1038/cr.2012.30}.
\newblock URL \url{http://www.nature.com/articles/cr201230}.

\bibitem[Ritchie et~al.(2015)Ritchie, Phipson, Wu, Hu, Law, Shi, and Smyth]{ritchie2015limma}
Matthew~E. Ritchie, Belinda Phipson, Di~Wu, Yifang Hu, Charity~W. Law, Wei Shi, and Gordon~K. Smyth.
\newblock Limma powers differential expression analyses for rna-sequencing and microarray studies.
\newblock \emph{Nucleic Acids Research}, 43\penalty0 (7):\penalty0 e47, 1 2015.
\newblock ISSN 13624962.
\newblock \doi{10.1093/nar/gkv007}.
\newblock URL \url{https://academic.oup.com/nar/article/43/7/e47/2414268}.

\bibitem[Robinson et~al.(2010)Robinson, McCarthy, and Smyth]{robinson2010edger}
M.~D. Robinson, D.~J. McCarthy, and G.~K. Smyth.
\newblock edger: a bioconductor package for differential expression analysis of digital gene expression data.
\newblock \emph{Bioinformatics}, 26\penalty0 (1):\penalty0 139--140, 1 2010.
\newblock ISSN 1367-4803.
\newblock \doi{10.1093/bioinformatics/btp616}.
\newblock URL \url{https://academic.oup.com/bioinformatics/article-lookup/doi/10.1093/bioinformatics/btp616}.

\bibitem[Tarazona et~al.(2011)Tarazona, Garc{\'{i}}a-Alcalde, Dopazo, Ferrer, and Conesa]{tarazona2011differential}
Sonia Tarazona, Fernando Garc{\'{i}}a-Alcalde, Joaquín Dopazo, Alberto Ferrer, and Ana Conesa.
\newblock Differential expression in rna-seq: A matter of depth.
\newblock \emph{Genome Research}, 21\penalty0 (12):\penalty0 2213--2223, 12 2011.
\newblock ISSN 1088-9051.
\newblock \doi{10.1101/gr.124321.111}.
\newblock URL \url{http://genome.cshlp.org/lookup/doi/10.1101/gr.124321.111}.

\bibitem[Trapnell et~al.(2009)Trapnell, Pachter, and Salzberg]{trapnell2009tophat}
Cole Trapnell, Lior Pachter, and Steven~L. Salzberg.
\newblock Tophat: discovering splice junctions with rna-seq.
\newblock \emph{Bioinformatics}, 25\penalty0 (9):\penalty0 1105--1111, 5 2009.
\newblock ISSN 1460-2059.
\newblock \doi{10.1093/bioinformatics/btp120}.
\newblock URL \url{https://academic.oup.com/bioinformatics/article-lookup/doi/10.1093/bioinformatics/btp120}.

\bibitem[Wickham(2015)]{Wickman2015Livros}
Hadley Wickham.
\newblock \emph{R Packages: Organize, Test, Document, and Share Your Code}.
\newblock O'Reilly Media, 1 edition, 4 2015.
\newblock ISBN 9781491910566.
\newblock URL \url{https://books.google.com.br/books?hl=pt-BR&lr=&id=DqSxBwAAQBAJ&oi=fnd&pg=PR3&dq=r+packages+test&ots=ao-8D0UMDh&sig=ieGPZ53bCXGEwIMvQjvxHHgq6JY#v=onepage&q=r%20packages%20test&f=false}.

\bibitem[Wickham et~al.(2022)Wickham, Hester, Chang, and Bryan]{devtools2022}
Hadley Wickham, Jim Hester, Winston Chang, and Jennifer Bryan.
\newblock \emph{devtools: Tools to Make Developing R Packages Easier}, 2022.
\newblock https://devtools.r-lib.org/, https://github.com/r-lib/devtools.

\bibitem[Wickham et~al.(2024)Wickham, Danenberg, Csárdi, and Eugster]{roxygen2024}
Hadley Wickham, Peter Danenberg, Gábor Csárdi, and Manuel Eugster.
\newblock \emph{roxygen2: In-Line Documentation for R}, 2024.
\newblock URL \url{https://roxygen2.r-lib.org/}.
\newblock R package version 7.3.2, https://github.com/r-lib/roxygen2.

\end{thebibliography}
\end{document}